\documentclass[aps,prl,twocolumn,showpacs,superscriptaddress]{revtex4-1}
\usepackage{amsmath}
\usepackage{epsfig}
\usepackage{color}
\usepackage{endnotes}
\usepackage{natbib}
\usepackage[colorlinks,breaklinks]{hyperref}
\usepackage{nicefrac}
\usepackage{bm}
\usepackage{dcolumn}
\usepackage{graphics}
\usepackage{amsmath}
\usepackage{amssymb}
\usepackage{color}
\usepackage{changes}
\newcommand{\bra}{\langle}
\newcommand{\ket}{\rangle}

\newcommand{\bs}[1]{\ensuremath{\boldsymbol{#1}}}

\newcommand{\be}{\begin{equation}}
\newcommand{\ee}{\end{equation}}
\newcommand{\bea}{\begin{align}}
\newcommand{\eea}{\end{align}}
\newcommand{\beqa}{\begin{eqnarray}}
\newcommand{\eeqa}{\end{eqnarray}}

\begin{document}

\title{Experimental evaluation of the nuclear neutron-proton contact}

\author{Ronen Weiss}
\affiliation{The Racah Institute of Physics, The Hebrew University, 
             Jerusalem, Israel}
\author{Betzalel Bazak}
\affiliation{The Racah Institute of Physics, The Hebrew University, 
             Jerusalem, Israel}
\author{Nir Barnea}
\email{nir@phys.huji.ac.il}
\affiliation{The Racah Institute of Physics, The Hebrew University, 
             Jerusalem, Israel}

\date{\today}

\begin{abstract} 
The nuclear neutron-proton contact is 
introduced, generalizing Tan's work, and
evaluated from medium energy 
nuclear photodisintegration experiments. To this end we reformulate the 
quasi-deuteron model of nuclear photodisintegration and establish the bridge 
between the Levinger constant and the contact.
Using experimental evaluations of Levinger's constant we extract the value of 
the neutron-proton contact in finite nuclei and 
in symmetric nuclear matter.
Assuming isospin symmetry we propose to evaluate the neutron-neutron contact 
through measurement of photonuclear spin correlated 
neutron-proton pairs.
\end{abstract}


\maketitle
{\it Introduction --}
Considering a system of two-component fermions interacting via a short range  
interaction, Tan \cite{Tan08,Bra12} has established a series of relations 
between the amplitude of the high-momentum tail of the momentum distribution 
$n_{\sigma}(k)$, where $\sigma$ is the spin, and the properties of the system, 
such as the energy, pair correlations and pressure.  
These relations, commonly known as the ``Tan relations'',
are expressed through a new variable the ``Contact''  
 $C=\lim_{k\rightarrow\infty}k^4 n_{\sigma}(k)$. 
The contact, being a state variable
depends on the density of the system 
(usually expressed through the Fermi momentum $k_F$), 
its temperature, composition,
and its thermodynamic state. 
The Tan relations are universal, they hold for few-body as well as 
for many-body systems, for ground state
and for finite temperature, for normal state but also for superfluid state.
Their validity range depends on the interparticle distance 
$d\propto 1/k_F$ and the magnitude of the scattering length
being both much larger than the potential range, usually
characterized by the effective range $r_\mathrm{eff}$.

The theoretical discovery of the Tan relations has led to a concentrated 
experimental effort to measure and verify them in ultracold atomic systems, 
where the scattering length as well as the density can be controlled. 
These efforts have led to experimental verification of Tan's relations in
fermionic $^{40}$K \cite{SteGaeDra10,SagDraPau12} and $^6$Li 
\cite{ParStrKam05,WerTarCas09,KuhHuLiu10} systems.
It was also found that the measured value of the contact, as function 
of $(k_F a)^{-1}$ along the BCS-BEC crossover, is in accordance with the 
theoretical predictions of \cite{WerTarCas09}.

In this manuscript we focus on nuclear systems.
Generalizing Tan's work, we introduce the nuclear contacts and present
an experimental evaluation of the neutron-proton contact in 
finite nuclei and also in symmetric nuclear 
matter. To this end we relate the contact to medium energy photonuclear 
cross-section and utilize available experimental data. 
In ultracold atomic physics the ratio between the interparticle
distance $1/k_F$, the scattering length $a$, and $r_\mathrm{eff}$ can be 
controlled in such a way as to ensure that the $a \gg r_\mathrm{eff}$ and 
$k_F r_\mathrm{eff}\ll 1$. 
The nuclear two--body scattering length is about 5.38 fm when the two nucleons 
are in the $^3S_1$ state and about -20 fm when they are in the $^1S_0$ state. 
These scattering lengths are denoted by $a_t$ for the $^3S_1$ channel, 
and $a_s$ for the $^1S_0$ channel.
The average interparticle distance in the atomic
nucleus is about $2.4\; \mathrm{fm}$. This number can be deduced from the 
empirical nuclear charged radius formula $R_c\approx 1.2 A^{1/3}\;\mathrm{fm}$. 
The long range part of the 
nuclear potential is governed by the pion exchange Yukawa force with 
characteristic length of 
$\mu^{-1}=\hbar/m_{\pi}c\approx 1.4 \;\mathrm{fm}$. 
Therefore in contrast with atomic physics, in nuclear physics the demand 
$k_Fr_\mathrm{eff}\ll 1$ can at best replaced by $\mu d > 1$ which holds for 
interparticle distance of about 2 fm. 

For photons in the energy range $\hbar\omega = 100-200 \;\mathrm{MeV}$, 
corresponding to the wave number $k \approx 0.5-1 \;\mathrm{fm}^{-1}$,
the deuteron photoabsorption cross-section is dominated by the leading 
electric dipole $E1$ and magnetic dipole $M1$ transitions 
\cite{Arenhovel91}. The nuclear photo effect 
at these energies is dominated by 
the quasi-deuteron process first proposed by Levinger more than 60 years ago 
\cite{Lev51}. In the quasi-deuteron picture the photonuclear
reaction mechanism goes 
through an absorption of the photon by a correlated proton-neutron ($pn$)
pair being close to each other, followed by an emission  
of the $pn$ pair back to back, flying without further interaction with the 
remaining nucleons. The resulting photonuclear cross-section of a nucleus 
composed of $Z$ protons and $N$ neutrons, $A=N+Z$, is therefore 
proportional to the deuteron cross-section $\sigma_d$,
\be \label{Levinger}
\sigma_A=L\frac{NZ}{A}\sigma_d 
\ee
with $L\approx 6$ being the Levinger constant. 
In the decades following Levinger's original work there were
few compilations of the photonuclear data and systematic evaluations of the 
Levinger constant (see {\it e.g.} \cite{TavTer92} and references therein).
It was also found that 2--body short range correlations captured so well
by the quasi-deuteron model plays an important role in analyzing hard nuclear 
electron scattering experiments, 
see {\it e.g.} \cite{LO90,FS88,WirSchPie14}.
Moreover, Levinger's picture has got a remarkable experimental support  
when high momentum, correlated, $pn$ pairs flying back to back where measured 
in proton and electron scattering on carbon \cite{Piasetzky06,Subedi08}
and other nuclei \cite{ArrHigRos12,Fom12}.

Already from the pictorial description of the quasi-deuteron model one can sense
the underlying connection between Tan's contact and Levinger's constant. 
In the following we shall define the various nuclear contacts associated with 
the permissible two-nucleon $s$-wave states. 
Utilizing these contacts we shall rederive the quasi-deuteron model,
establishing the desired connection. 
As will be evident later, the nuclear contacts can be evaluated from
either spin independent transitions or from experiments on spherical nuclei. 
Consequently  
we shall concentrate on the $E1$ transition cross-section, which in principle 
can be extracted from the angular distribution of the emitted $pn$ pair.
The experimental evaluation of Levinger's constant \cite{TavTer92} is than
used to extract the proton-neutron contact.

{\it The Contact in Nuclear Systems --}
When two particles interacting via short range force
approach each other, the many-body wave function can be factorized into a 
product of an asymptotic pair wave function $\varphi_{ij}(\bs r_{ij})$,
where $\bs r_{ij}=\bs{r}_i-\bs{r}_j$,
and a function $A_{ij}$, also called the regular part of $\Psi$,
describing the residual $A-2$ particle system
and the pair's center of mass (CM) $\bs{R}_{ij}=(\bs{r}_i+\bs{r}_j)/2$
motion \cite{Tan08,WerCas12},
\be \label{wf}
  \Psi \xrightarrow[r_{ij}\rightarrow 0]{}\varphi_{ij}(\bs{r}_{ij})
           A_{ij}(\bs{R}_{ij},\{\bs r_k\}_{k\neq i,j})\;.
\ee
Due to the suppression of higher partial waves, the pair wave function
will be predominantly an $s$-wave.
In particular, in the zero-range model \cite{zerorange}, where the action of an interacting 
particle pair with scattering length $a$ on the many-body wave-function 
$\Psi$ is represented through the boundary condition 
$
 \left[{\partial \log (r_{ij}\Psi)}/{\partial r_{ij}}\right]_{r_{ij}=0}=-{1}/{a}
$\;,
the low energy asymptotic pair wave function takes a particularly simple form
$\varphi_{ij}=\left(1/r_{ij}-1/a\right)$.

The contact $C$ represents the probability of finding a correlated
 pair within the system, and can be expressed as \cite{Tan08,WerCas12}
\be\label{contact_generic}
   C=16\pi^2 \sum_{i < j} \bra A_{ij}| A_{ij}\ket
\ee
where 
\begin{align}
   \bra A_{ij}|  A_{ij}\ket & = 
    \int  \prod_{k\neq i,j} d\bs{r}_{k} \,d\bs{R}_{ij} \,
    \\ \nonumber & \times
        A_{ij}^{\dagger}\left(\bs{R}_{ij},\{\bs{r}_{k}\}_{k\neq i,j}\right)
        \cdot
        A_{ij}\left(\bs{R}_{ij},\{\bs{r}_{k}\}_{k\neq i,j}\right)\;
\end{align}
is independent of the particular form of the asymptotic pair wave function 
$\varphi_{ij}$. The Pauli principle implies that $A_{ij}=0$ if the two particles
are in the same internal state.

Generalizing this formalism to nuclear systems, 
the pair can be in more than one configuration, and we have to 
consider six possible pairs
$P=\{p\!\uparrow\!p\!\downarrow, 
   n\!\uparrow\!n\!\downarrow,
   p\!\uparrow\!n\!\downarrow,
   p\!\uparrow\!n\!\uparrow,
   p\!\downarrow\!n\!\uparrow,
   p\!\downarrow\!n\!\downarrow\}$.
In this representation we can define a contact $C_{P}$ for each pair $P$. 
These contacts are proportional to 
the diagonal elements of the overlap matrix $\bra A_{ij}^P|A_{ij}^{P'}\ket$.
In nuclear physics it is more natural, however, to employ a spin-isospin
basis that diagonalize the overlap matrix.
Furthermore, assuming now spin symmetry for the residual function norm
$\bra A_{ij}^P| A_{ij}^P\ket $
we have to consider only 
four contacts associated with the pairs
$P=\{(pp)_{S=0},(nn)_{S=0},(np)_{S=0},(np)_{S=1}\}$. Taking into account that the 
coulomb force as well as other isospin symmetry-breaking terms are 
negligible at short distances, the number of independent nuclear contacts in 
symmetric nuclei $(N=Z)$ can be further reduced to 
only two, corresponding to the two-body spin-isospin configurations
$\{|S=0,T=1\ket,|S=1,T=0\ket\}$. 
In the quasi-deuteron mechanism described above only correlated $pn$ pairs 
play any role, thus we need to consider only the two nuclear contacts 
$\{C_{s},C_{t}\}$ corresponding to the spin singlet and spin triplet states.

In bosonic systems the high momentum tail of the momentum distribution
contains a $1/k^5$ correction due to three body contact \cite{BraKanPla11}. 
We note that such correction is to be expected in nuclear systems, 
as three nucleon coalescence is not forbidden by the Pauli exclusion principle. 
Studying 3-body effects on the nuclear photoabsorption cross-section and 
consequently estimating the nuclear 3-body contact is an important task. 
Nevertheless in the current letter we focus on the leading 2-body effect.

{\it The Quasi-Deuteron model in the zero range approximation -- }
In the following we will utilize the zero range approximation 
to relate the contact
to the quasi-deutron model. 
This model allows a clear and simple derivation.
We note however,
that it can be easily generalized to include more realistic wave 
functions.

In the leading $E1$ approximation, the total photo absorption cross section of 
a nucleus is given by 
\begin{equation}\label{cross_section_A}
\sigma_A(\omega)=4\pi^{2}\alpha\hbar\omega R(\omega)\,,
\end{equation}
\noindent where $\alpha$ is the fine structure constant,
\begin{equation} \label{1}
  R(\omega )=\bar{\sum_i} \sum_{f}\left| \bra \Psi _{f}\right| \bs{\epsilon}\cdot
   \hat{\bs D} \left| 
\Psi _{0}\ket \right| ^{2}\delta (E_{f}-E_{0}-\hbar\omega) 
\end{equation}
is the response function, $\hat{\bs D}$ is the unretarded dipole operator
$\hat{\bs D}=\sum_{i=1}^{A}\frac{1+\tau^{3}_{i}}{2}\bs r_{i}$, 
and ${\bs \epsilon}$ is the photon's polarization vector. 
The wave functions of
the initial (ground) state and of the final state are denoted by 
$\left| \Psi_{0/f} \right\ket$ and the energies by $E_{0/f}$, respectively. 
The operator $\tau^{3}_{i}$ is the third components of the $i$-th nucleon 
isospin operator. 
The response functions includes a sum over the final states $\sum_f$ that
becomes an integration in the limit of infinite volume, and an average over
the initial states which amounts to an average over the 
magnetic projection of the ground state, $\bar\sum_i=1/(2J_0+1)\sum_{M_0}$.  

For inverse photon wave number somewhat shorter than the average interparticle 
distance ($kd>1$),
the reaction cross-section goes via a nucleon pair that absorbs the photon.
The $E1$ nature of the process implies that the pair must be a neutron-proton 
pair since proton-proton pair posses no dipole moment. 
Utilizing the zero-range approximation,
\begin{align}
\label{eq:asymptotic form}
\Psi_0 \xrightarrow[r_{pn}\rightarrow 0]{} &
  \sum_{P}
   \left(\frac{1}{r_{pn}}-\frac{1}{a_{P}}\right)
   A^P_{pn}\left(\bs{R}_{pn},\{\bs{r}_{j}\}_{j\neq p,n}\right) \cr &
  +O(r_{pn}),
\end{align}
where
$A^P_{pn}=
   \sum_{J_{A-2}}\left[\chi_{P}\otimes 
       A_{P}^{J_{A-2}}(\bs{R}_{pn},\{\bs{r}_{j}\}_{j\neq p,n})\right]^{J_0 M_0}.$
Here the notation $pn$ stands for any proton-neutron pair, 
whose spinors are coupled into a spin state $\chi_P$ with total spin $S=0,1$,
and the corresponding scattering length $a_P$.
The sum over the angular momentum of the remaining $A-2$ nucleons $J_{A-2}$ 
extends over all possible configurations that coupled to $S$ yield 
the ground state's total angular momentum quantum numbers $J_0,M_0$.

Turning now to the final state, we consider a reaction mechanism 
where the photon is absorbed by a proton $p$ that is emitted with large momentum
$\mathbf{k}_{p}$. For high photon energies this process is fast
enough and interaction between the emitted proton and rest of the nucleus can be
neglected, that is the Born approximation.
Hence, momentum conservation implies that another particle must be emitted. As
pointed out by Levinger \cite{Lev51}, this particle must be a neutron 
$n$ emitted with momentum $\mathbf{k}_{n}$, such as 
$\mathbf{k}_{n}\approx-\mathbf{k}_{p}\equiv\mathbf{k}$.
The relative momentum of the emitted particles is 
$\frac{\mathbf{k}_{n}-\mathbf{k}_{p}}{2}=\frac{2\mathbf{k}_{n}}{2}=\mathbf{k}$,
and they can form either an $S=0$ or an $S=1$ spin states.
Assuming that the residual $A-2$ particles wave function is frozen throughout 
this process, the final state wave function for an outgoing spin $S$ pair is
given by
\be \label{eq:final state}
\Psi_{f}^P={\cal N}_P
      \hat{\cal A}\left\{\frac{1}{\sqrt{\Omega}}e^{-i\mathbf{k}\cdot\mathbf{r}_{pn}}
      A^P_{pn}(\bs{R}_{pn},\{\bs{r}_{j}\}_{j\neq p,n})\right\}\;,
\ee
where ${\cal N}_P$ is a normalization factor, 
the wave function is normalized in a box of volume $\Omega$, and
$ \hat{\cal A}=\big(1-\sum_{p'\neq p}(p,p')\big)
             \big(1-\sum_{n'\neq n}(n,n')\big) $
is the proton-neutron antisymmetrization operator with $(i,j)$ the 
transposition operator. 
The sums over $p', n'$ extends over all protons and neutrons in the system but 
$p, n$.
As $A^P_{pn}(\bs{R}_{pn},\{\bs{r}_{j}\}_{j\neq p,n})$ is antisymmetric under 
permutation of all identical particles but the pair $pn$, $\Psi_{f}^P$ is
antisymmetric under proton permutations and under neutron permutations.

Utilizing now permutational symmetry, the nuclear  
neutron-proton contacts $C_P=\{C_s,C_t\}$ are given by 
\be
 C_P=16\pi^2 N Z \bra A^P_{pn}|A^P_{pn}\ket \;.
\ee
Therefore, the normalization factor is given by
${\cal N}_P=\frac{1}{\sqrt{NZ}}\frac{1}{\sqrt{\bra A^P_{pn}|A^P_{pn}\ket}}
  =\frac{4\pi}{\sqrt{C_P}}$.
Considering now the transition matrix element we see that
\begin{align}
\bra \Psi_{f}^{P}|&\bs{\epsilon}\cdot\hat{\bs D}|\Psi_{0}\ket 
    = N Z {\cal N}_P 
    \int \prod_{k}d\bs{r}_{k} \\ \nonumber
    & \times
    \frac{1}{\sqrt{\Omega}}e^{i\mathbf{k}\cdot\mathbf{r}_{pn}}
    A_{pn}^{P\dagger}\left(\mathbf{R}_{pn},\{\mathbf{r}_{j}\}_{j\neq pn}\right)
    \left(\bs{\epsilon}\cdot\hat{\bs D}\right)\Psi_{0}
\end{align}
where we have used the fact that $\hat{\cal A}\Psi_{0}=NZ\Psi_{0}$.
Due to the orthogonality of the initial and final states the transition 
matrix element vanishes unless the photon acts on the outgoing $pn$ pair.
Since the momentum $\mathbf{k}$ is large, the only
significant contribution to the integral comes from the asymptotic
$r_{pn}\rightarrow 0$, where $\Psi_{0}$ diverges, and therefore
the integration over $r_{pn}$ hereafter can be limited to a small neighborhood 
of the origin $\Omega_0$. See Supplemental Material  
\cite{supplemental} for more details.
Hence,
\begin{align}\label{D_me2}
\bra \Psi_{f}^{P}|&\bs{\epsilon}\cdot\hat{\bs D}|\Psi_{0}\ket 
    = N Z {\cal N}_P \sum_{P'} \bra A^P_{pn} | A^{P'}_{pn} \ket 
     \\ \nonumber & \times \int_{\Omega_0} d\bs{r}_{pn}
    \frac{1}{\sqrt{\Omega}}e^{i\mathbf{k}\cdot\mathbf{r}_{pn}}
    \left(\bs{\epsilon}\cdot\hat{\bs D}_{pn}\right)
    \left(\frac{1}{r_{pn}}-\frac{1}{a_{P'}}\right),
\end{align}
where $\hat{\bs{D}}_{pn}=\mathbf{r}_{p}\simeq\frac{\mathbf{r}_{pn}}{2}$
neglecting the pair's CM motion.
Most of the photon energy is delivered to the relative motion whereas the 
photon's momentum is translated into the CM motion, thus the energy fraction 
associated with the CM coordinate $\bs{R}_{pn}$ is $\hbar\omega/4 Mc^2$ which 
amounts to only few percents for the photon energies under consideration. 
We can therefore safely neglect the pair's recoil. We note that the matrix 
element (\ref{D_me2}) is independent of $M_0$ thus $\bar\sum_i=1$ in 
Eq. (\ref{1}).
For the $E1$ operator or for any spin scalar operator the orthogonality 
of the different two-body spin functions included in $A_{P}$ ensures
that the spin state of the $pn$ pair is unaltered throughout the process, 
i.e. $P'=P$.
For spherical $J_0=0$ nuclei this important result 
holds for any one-body nuclear current operator since the different singlet and
triplet spin states must be coupled to spectator functions $A_P^{J_{A-2}}$ 
with $J_{A-2}=0,1$ respectively. These spectator functions are orthogonal 
and therefore there is no interference between the different
$pn$ spin states.
Utilizing these results we can rewrite the transition matrix element in the form
\begin{align}\label{me_A}
   \bra\Psi_{f}^{P}|\bs{\epsilon}&\cdot\hat{\bs D}|\Psi_{0}\ket
    = \\ \nonumber
    &\frac{\sqrt{C_P}}{4\pi}
    \int_{\Omega_0} d\bs{r}_{pn}\frac{1}{\sqrt{\Omega}}e^{i\mathbf{k}\cdot\mathbf{r}_{pn}}     
    \bs{\epsilon}\cdot\hat{\bs D}_{pn}
    \left(\frac{1}{r_{pn}}-\frac{1}{a_{P}}\right) \;.
\end{align}
The matrix element (\ref{me_A}) looks very much as the deuteron's 
photoabsorption transition matrix element.
To make this comparison complete, let us consider the deuteron's 
photoabsorption reaction. The deuteron is a bound proton-neutron pair 
with angular momentum $J=1$. In the zero 
range approximation, the deuteron wave function is assumed to be a pure 
$s$-wave, spin triplet state, that takes the form
\begin{equation}
\psi_{d,0}
    = \frac{1}{\sqrt{2\pi a_{t}}} \frac{e^{-r_{pn}/a_{t}}}{r_{pn}}
     \xrightarrow[r_{pn}\rightarrow 0]{}
     \frac{1}{\sqrt{2\pi a_{t}}}\left(\frac{1}{r_{pn}}-\frac{1}{a_{t}}\right).
\end{equation}
In the Born approximation, and neglecting the CM recoil, the deuteron's final 
state wave function is given by
\begin{equation}
\psi_{d,f}=\frac{1}{\sqrt{\Omega}}e^{-i\mathbf{k}\cdot\mathbf{r}_{pn}}.
\end{equation}
Hence, 
\begin{align}\label{me_d} 
  \bra \psi_{d,f}|& \bs{\epsilon}\cdot\hat{\bs D}|\psi_{d,0}\ket 
  = \\ \nonumber &
  \int d\bs{r}_{pn} \frac{1}{\sqrt{\Omega}}e^{i\mathbf{k}\cdot\mathbf{r}_{pn}}
                   \bs{\epsilon}\cdot\hat{\bs D}_{pn}
                   \frac{1}{\sqrt{2\pi a_{t}}} \frac{e^{-r_{pn}/a_{t}}}{r_{pn}}.
\end{align}
Analyzing Eqs. (\ref{me_A}) and (\ref{me_d})
we note that in the high momentum limit $\bs{k}\longrightarrow\infty$ 
the main contribution to the transition matrix-element emerge
from $\Omega_0$ the neighborhood of the origin $r_{pn}=0$. 
There, the $np$ pair wave function takes the asymptotic form 
$\left(1/{r_{pn}}-{1}/{a_{t}}\right)\approx 1/r_{pn}$.
Utilizing this approximation and
comparing Eqs. (\ref{me_A}) and (\ref{me_d}) we can conclude that
\be
  \bra \Psi_f^P | \bs{\epsilon}\cdot\hat{\bs D} | \Psi_0 \ket \approx
   \sqrt{\frac{C_P a_t}{8\pi}}
   \bra \psi_{d,f}|\bs{\epsilon}\cdot\hat{\bs D}|\psi_{d,0}\ket
\ee
Substituting this result into Eq. (\ref{cross_section_A}) and summing over
the possible final state spin configurations 
we get 
\be\label{WBB}
\sigma_A(\omega)= \frac{a_t}{8\pi}(C_s+C_t)\sigma_d(\omega)
\ee
where $\sigma_d(\omega)$ is the deuteron photonuclear cross-section.
Comparing this result with the celebrated Levinger formula, 
Eq. (\ref{Levinger}),
we see that the Levinger constant $L$ can be directly expressed through 
the nuclear contacts $C_P$,
\be \label{Lev}
   L= \frac{a_t}{8\pi}\frac{A}{N Z}(C_s+C_t)\;.
\ee
The Levinger constant was explored and evaluated in various photonuclear 
experiments. Using this data
the averaged nuclear $pn$ contact $\bar C_{pn}\equiv (C_s+C_t)/2$ can be 
evaluated. 

Before we proceed to actual evaluation of the nuclear contact few comments 
are in place: 
(i) In our derivation we have utilized the zero range approximation. 
The validity of this approximation is at best questionable for finite nuclei.
Nonetheless, the derivation can be generalized to any short range pair wave 
function given that it is unique across the nuclear chart.
In this case, the resulting many-body contact 
should be expressed in terms of the deuteron contact, i.e. $8\pi/a_t$ is to 
be replaced by the deutron contact in Eqs. (\ref{WBB}) and (\ref{Lev}).
(ii) Although we have only considered the 
dipole response, our main result (\ref{WBB}) holds in spherical $J_0=0$ 
nuclei for any multipole and any one-body current operator. 
For $J_0\neq 0$ nuclei, Eq. (\ref{WBB}) holds for any spin independent 
one-body operator.
(iii) If instead of measuring the total photoabsorption cross-section one 
measures the cross-section $\sigma_A^{\uparrow\!\uparrow}$ for the parallel spin 
reaction $\gamma+^A\!X\longrightarrow ^{A-2}\!Y + p\!\uparrow\!n\!\uparrow$,
we would obtain 
\be
\sigma_A^{\uparrow\!\uparrow}(\omega)= \frac{1}{3}\frac{a_t}{8\pi}C_t\sigma_d(\omega)
\ee
for spherical $J_0=0$ nuclei.
Thus, measuring the spin correlated photonuclear cross-section
would enable the separation of the two nuclear contacts $C_s$ and $C_t$.

\begin{figure}\begin{center}
\includegraphics[width=8.6 cm]{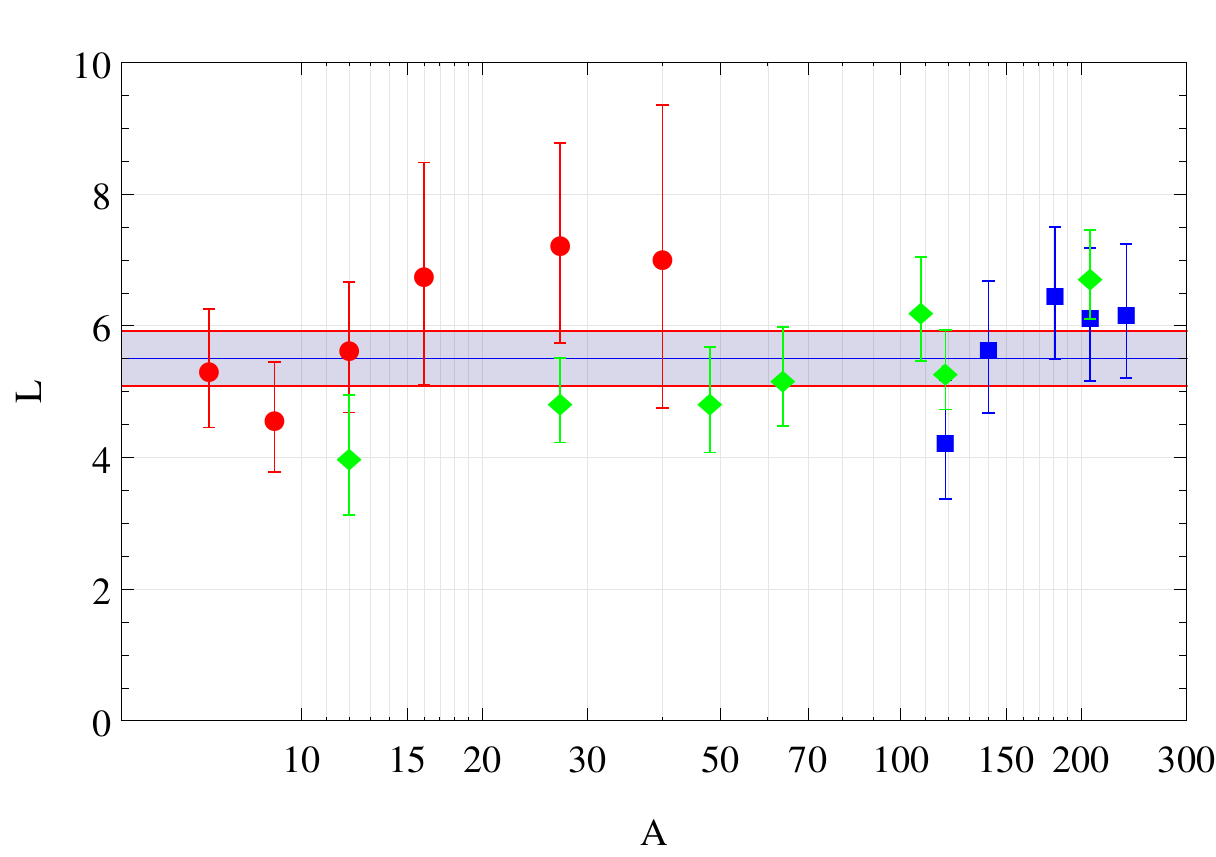}
\caption{\label{LevP} (Color online) 
Levinger constant values evaluated from photonuclear experiments. 
Red circles are based on Ref. \cite{AhrBorCzo75}, 
blue squares are based on Ref. \cite{Lep78},
and green diamonds are based on Ref. \cite{Sti84}.
Adapted with changes from \cite{TavTer92}.
The blue line shows the averaged value $L=5.50$, and the gray band its error.}
\end{center}\end{figure}

{\it Experimental evaluation of the nuclear neutron-proton contact -- }
At this point we would like to extract the nuclear neutron-proton contact from 
the experimental photonuclear data. 
To this end we use an analysis of the Levinger constant made by 
Terranova {\it et al.} \cite{TavTer92}, who evaluated the 
Levinger constant $L$ for 14 nuclei along the periodic table, from lithium to 
uranium, using various photonuclear experiments
\cite{AhrBorCzo75,Sti84,Lep78,Hom83}. In order to include low energy
data Terranova {\it et al.} have used in their analysis the modified 
quasi-deuteron model \cite{MQD}, 
taking into account the Pauli blocking. 
For high photon energies this is a small correction.
The evaluated Levinger constant, presented in Fig. \ref{LevP}, 
seems to be constant along the nuclear chart, with an averaged value
of $L \approx 5.50 \pm 0.21\;(1\sigma)$.
Using this result we can estimate the average $pn$ contact for symmetric nuclei 
and nuclear matter, namely 
\be\label{cbarpn}
  \frac{\bar C_{pn}}{k_F A}=\frac{\pi}{k_Fa_t}\left(5.50\pm 0.21\right)
       \approx 2.55 \pm 0.10  \;.
\ee
Here in the last equality we have used the relation $1/k_Fa_t \approx 0.15$
valid on the average for large nuclei. We note that the quoted error in 
Eq. (\ref{cbarpn}) refers only to statistical errors, and not to systematic 
errors associated with our model assumptions. 

\begin{figure}\begin{center}
\includegraphics[width=8.6 cm]{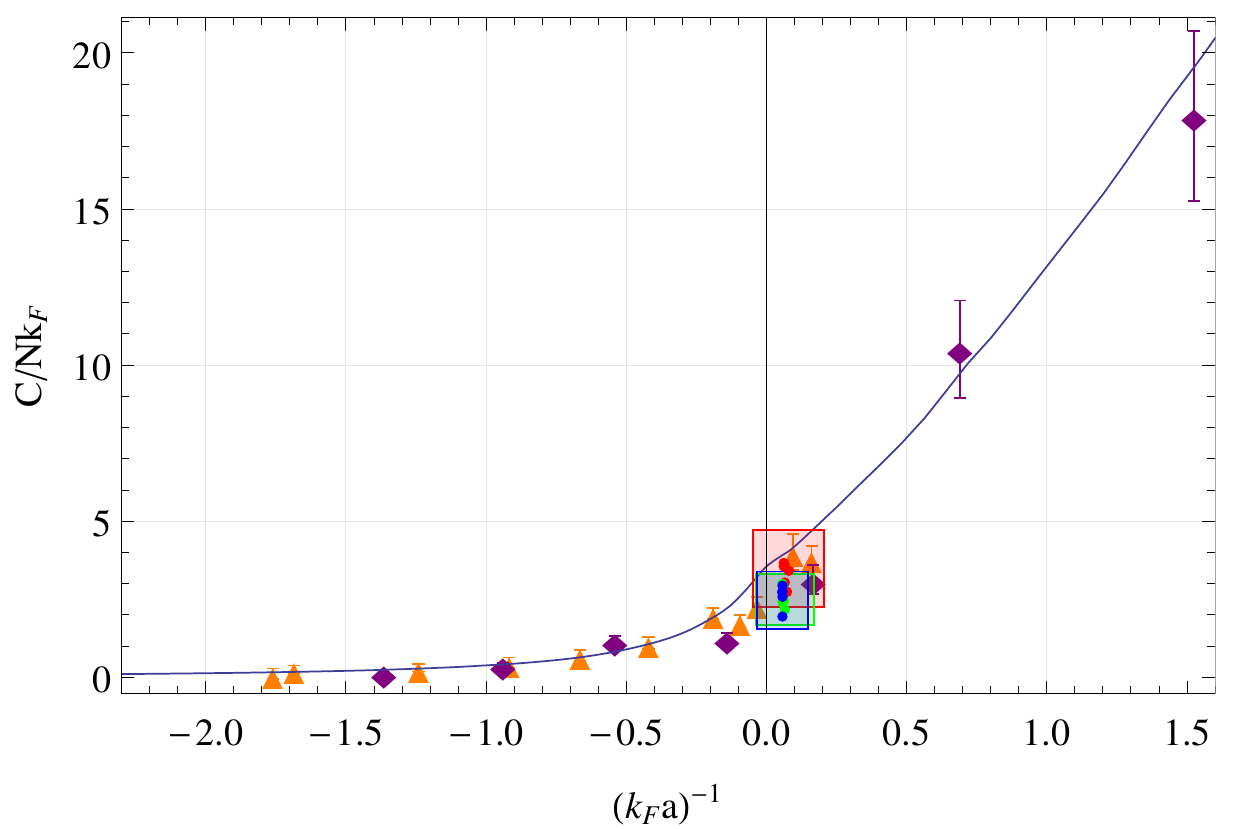}
\caption{\label{Cpn} (Color online) 
The universal contact per particle (in units of $k_F$), 
measured in ultracold atoms, compared to the 
averaged $pn$
contact evaluated here from photonuclear experiments.
Orange triangles - measurement with ultracold  $^{40}$K atoms \cite{SteGaeDra10}, 
purple diamonds - ultracold $^6$Li atoms \cite{ParStrKam05,WerTarCas09}.
Red, blue and green dots - nuclear contact based on 
photonuclear experiments of \cite{AhrBorCzo75}, \cite{Lep78}, and
\cite{Sti84}. The square's bounds represent the experimental error
for the different data sets.
The line is the theoretical prediction of Ref. \cite{WerTarCas09}.}
\end{center}
\end{figure}

As mentioned above, the contact was measured for a universal Fermi gas
along the BCS-BEC crossover, i.e. as a function of the dimensionless parameter
$k_F a$.
In order to compare the nuclear $pn$ contact to the universal Fermi gas results
we estimated $k_F$ for each nuclei using the rms charge radius, evaluated by 
Brown {\it et al.} \cite{BroBroHod84}.
In Fig. \ref{Cpn} we present the universal Fermi gas contact 
measured with $^{40}$K atoms \cite{SteGaeDra10}, and
 $^6$Li atoms \cite{ParStrKam05,WerTarCas09}, the  
theoretical prediction of Ref. \cite{WerTarCas09}, and
the average nuclear $pn$ contact evaluated for each nucleus individually.
For the nuclear scattering length we have used $1/a=1/2(1/a_s+1/a_t)$,
with error bar that corresponds to the difference between the singlet and 
triplet scattering lengths.
Inspecting the figure it is noteworthy that although nuclei are far 
from being a universal Fermi gas, the nuclear contact 
falls in line with that of a universal Fermi gas.

{\it Summary -- } 
Summing up, 
rederiving the quasi-deuteron model in the zero range approximation we have
constructed a bridge between the contact $C$, recently introduced by Tan to
describe the properties of interacting Fermi systems, and nuclear systems. 
Doing so we have identified two contacts $C_s,\;C_t$, corresponding to spin 
singlet and spin triplet states, and have shown that the average $pn$
contact $\bar C_{pn}$ is proportional to Levinger's constant $L$. 
Using experimental estimates
for $L$ we have deduced the value of $\bar C_{pn}$.
We have found that the evaluated value of $\bar C_{pn}$
stands in good agreement with the universal contact measured in ultracold 
atomic experiments. This result hints towards the usefulness of Tan's relations
also in nuclear physics.
To separate between the singlet and triplet contacts we propose to measure the 
spin correlated emitted pairs in photonuclear experiment.

\begin{acknowledgments}
This work was supported by the Pazi fund. We thank O. Chen and E. Piasetzky 
for useful discussions, and O. A. P. Tavares and M. L. Terranova for 
their help retrieving the experimental data.
\end{acknowledgments}


\pagebreak
\widetext
\begin{center}
\textbf{\large Supplemental Materials: Experimental evaluation of the nuclear neutron-proton contact}
\end{center}
\setcounter{equation}{0}
\setcounter{figure}{0}
\setcounter{table}{0}
\setcounter{page}{1}
\makeatletter
\renewcommand{\theequation}{S\arabic{equation}}
\renewcommand{\thefigure}{S\arabic{figure}}
\renewcommand{\bibnumfmt}[1]{[S#1]}
\renewcommand{\citenumfont}[1]{S#1}

In our letter we have argued that in the large k limit the photodisintegration
matrix element is sensitive only to the most diverging part of the wave 
function, namely its behavior in short interparticle distances,
and therefore we limit our integrals, for example Eq. (11), 
to a small neighborhood of the origin $\Omega_0$. 
Here we present in details the calculation
of such integrals.

In the zero range approximation, the deuteron's photodisintegration 
cross section is proportional to the integral
\be
\int d\bs{r} e^{i\bs{k}\cdot\bs{r}}\bs{\epsilon}\cdot\bs{r}\frac{e^{-r/a}}{r}.\label{eq:exact_deuteron}
\ee
For $k\longrightarrow\infty$, the main contribution to this integral comes from 
the neighborhood of the origin $r=0$, because the fast-oscillating 
$e^{i\mathbf{k}\cdot\mathbf{r}}$ washes out anything but the most diverging part of the 
function. 
In this neighborhood $e^{-r/a}\approx1$, and the integral can be 
approximated by 
\be
\int_{\Omega_{0}} d\bs{r} e^{i\bs{k}\cdot\bs{r}}\bs{\epsilon}\cdot\bs{r}\frac{1}{r},
\label{eq:approx_deuteron}
\ee
where the integration is limited to a small neighborhood of the origin
$\Omega_{0}$ by some smooth cutoff function $f_R(r)$,
\be 
\int_{\Omega_0} d \bs r g(r) \equiv \int d\bs r g(r) f_R(r),
\ee
where $f_R(0)=1$.

First we note that similar approximation is used to prove the fundamental Tan 
relation \cite{Tan08}, namely
\be \label{Tan}
\lim_{k\rightarrow\infty} n_{\sigma}(k) =\frac{C}{k^4}.
\ee
In the two body case \cite{ComAlzLey09}, one has to Fourier transform the 
dimer wave function,
\be \label{exp}
\int d \bs r e^{i \bs k \cdot \bs r}\frac{e^{-r/a}}{r}= \frac{4\pi a^2}{a^2k^2+1}
\xrightarrow[ka\gg1]{} \frac{4\pi}{k^2}.
\ee
To show that this integral is dominated, in the large k limit, by its 
short-range behavior, let's approximate it by 
\be \label{OneOverR}
\int_{\Omega_0} d \bs r e^{-i \bs k \cdot \bs r}\frac{1}{r}
\ee
and use $f_R(r)=e^{-r/R}$ as a cutoff function.
It is clear that the resulting integral is equivalent to (\ref{exp}),
and therefore reproduce the right limit.
Using a Gaussian cutoff function 
$f_R(r)=e^{-(r/R)^2}$ one gets
\be
\int d\bs r e^{i\bs{k}\cdot\bs{r}} \frac{1}{r} e^{-(r/R)^2}=
\frac{4\pi R}{k}F\left(\frac{k R}{2}\right),
\ee
where $F(x)=e^{-x^{2}}\int_{0}^{x} dy e^{y^{2}}$ is the Dawson integral.
For large $x$, 
$F(x)=(2x)^{-1}+(4x)^{-3}+O(x^{-4})$
and in the limit $kR\gg1$ we obtain again $4\pi/k^2$.
In fact, we may conclude that 
any smooth cutoff function such
as $\exp(-r/R)$, $exp(-(r/R)^2)$, or $(1-\tanh((r-R)/h))/(1+\tanh(R/h))$
will reproduce this result in the 
high momentum limit.
In contrast, a sharp cutoff such as 
$f_R(r)=\Theta(R-r)$ will not work, because of the Gibbs phenomenon.

Note that here the same result can be achieved with $f_R(r)=1$, 
utilizing the relation $\Delta(1/r)=-4\pi\delta(\bs r)$. 

Now we go back to Eq. (\ref{eq:exact_deuteron}) and show that
the same approximation works there. First, operating with 
$\bs \epsilon \cdot \nabla_k$ on Eq. (\ref{exp}),
we get
\be
\int d\bs r e^{i\bs k \cdot \bs r}\bs \epsilon \cdot \bs r \frac{e^{-r/a}}{r}=
\frac{8\pi ia^{4}k}{\left(a^{2}k^{2}+1\right)^{2}}\bs \epsilon \cdot \hat{k}
\xrightarrow[ka\gg1]{ }
\frac{8\pi i\bs \epsilon \cdot \hat{k}}{k^{3}}.\label{eq:deuteron_large_k}
\ee

Once again a smooth cutoff function $f_R(r)=e^{-r/R}$ can be added to limit
the integral in Eq. (\ref{eq:approx_deuteron}) to a small neighborhood 
of the origin, yielding the same result for large $k$. We can also check
the use of a Gaussian cutoff,
\be
\int d\bs r e^{i\bs{k}\cdot\bs{r}}\bs{\epsilon}\cdot\bs r\frac{1}{r}e^{-(r/R)^2}
 = 2\pi i\frac{R}{k^2}\bs \epsilon\cdot\hat{k}\left(-kR+\left((kR)^2
+2\right)F\left(\frac{kR}{2}\right)\right) 
\xrightarrow[kR\gg1]{ }
\frac{8\pi i\bs \epsilon \cdot\hat{k}}{k^{3}}.
\ee
We may conclude that also in this case
any smooth cutoff function will reproduce the right  
high momentum limit.

In conclusion, we have shown here that indeed for $k\longrightarrow\infty$
the main contribution to the integrals in Eqs.  
(\ref{eq:exact_deuteron}) and (\ref{exp}) comes from a small 
neighborhood of the origin.
We have explained that it is important to limit the integrals by a smooth 
cutoff function, and we have checked it explicitly 
for exponential, Gaussian, and $\tanh$ cutoffs.

\end{document}